\documentclass{emulateapj}

\def\kms{km\,s$^{-1}$}
\def\teff{$T_{\rm eff}$}

\shorttitle{Double Degenerate NLTT~16249}
\shortauthors{Vennes et al.}

\begin{document}

\title{The 1.17-day orbit of the double-degenerate (DA+DQ) NLTT~16249 
\footnote{Based on observations made with ESO telescopes at the La Silla Paranal Observatory 
under programme ID 86.D-0562.}}

\author{S. Vennes$^{1,}$\altaffilmark{4}, A. Kawka$^{1,}$\altaffilmark{4}, S. J. O'Toole$^{2}$, J. R. Thorstensen$^{3}$ }
\affil{$^1$ Astronomick\'y \'ustav, Akademie v\v{e}d \v{C}esk\'e republiky, Fri\v{c}ova 298, CZ-251 65 Ond\v{r}ejov, Czech Republic}
\affil{$^2$ Australian Astronomical Observatory, PO Box 296, 1710 Epping NSW, Australia}
\affil{$^3$ Department of Physics and Astronomy, 6127 Wilder Laboratory, Dartmouth College, Hanover, NH 03755-352}


\altaffiltext{4}{Visiting Astronomer, Kitt Peak National Observatory, National Optical Astronomy Observatory, 
which is operated by the Association of Universities for Research in Astronomy (AURA) under cooperative 
agreement with the National Science Foundation.}

\begin{abstract}
New spectroscopic observations show that the double degenerate system NLTT~16249 is in a close 
orbit ($a=5.6\pm0.3\,R_\odot$) with a period of 1.17~d. The total mass of the system is 
estimated between 1.47 and 2.04\,$M_\odot$ but it is not expected to merge within
a Hubble time-scale ($t_{\rm merge}\approx 10^{11}$ yr). Vennes \& Kawka (2012, ApJ, 745, 
L12) originally identified the system because of the peculiar composite hydrogen (DA class) and molecular 
(C$_2$--DQ class--and CN) spectra and the new observations establish this system as the first DA plus DQ close double degenerate. Also, the DQ component was the first of its class to show
nitrogen dredged-up from the core in its atmosphere. The star may be viewed as the first known DQ descendant of the
born-again PG1159 stars. 
Alternatively, the presence of nitrogen may be the result of 
past interactions and truncated evolution in a close binary system.
\end{abstract}

\keywords{binaries: close -- stars: fundamental parameters --  white dwarfs}

\section{Introduction}

The high-proper motion star NLTT~16249 is a double degenerate system showing hydrogen
lines and molecular carbon and cyanogen bands in its optical spectrum \citep{ven2012}.
Although large radial velocity variations were noted by \citet{ven2012}, the orbital
parameters, period and separation, are yet to be determined.
The detection of photospheric nitrogen in the carbon-rich (DQ) component of this system was a first occurrence for
this class of objects and, according to the dredge-up scenario commonly applied to DQ
white dwarfs \citep{pel1986,mac1998}, it implied the presence of nitrogen in the white dwarf core.
The other component is a hydrogen-rich (DA) white dwarf with a mass above average but with a
luminosity similar to that of the DQ white dwarf.

The two characteristics of the system, a nitrogen-enriched DQ component and a likely close orbit, may
or may not be related. \citet{alt2005} established a clear evolutionary link between born-again stars
and a nitrogen/oxygen enrichment in PG1159 stars and their DQ descendents.
Members of the PG1159 class have a helium-rich surface with notable
enrichment in carbon, nitrogen and oxygen \citep{wer1991,dre1998,wer2006}. 
\citet{alt2005} argued that not all DQ stars followed the born-again path
because the carbon abundance is lower in some of these objects than would be expected 
following this evolutionary path. However, the 
$\log{\rm C/He}$ versus \teff\ trend established by \citet{duf2005} and \citet{koe2006} does appear to
follow the model predictions of \citet{alt2005}. Moreover, \citet{duf2005} noted two
discernible tracks with one at a
higher carbon abundance that was attributed to a thinning of the helium layer.
Therefore, data and born-again models may be compatible for the bulk of DQ white dwarfs. However, nitrogen is not detected in any
DQ stars other than in the DQ in NLTT~16249.

In this Letter, we present convincing evidence that the components of the binary NLTT~16249 are in a close orbit. Our new radial
velocity measurements (Section 2) and our original model atmosphere analysis \citep{ven2012}
help constrain the component properties and offer clues to the origin and evolutionary
prospect of the system (Section 3). We summarize and discuss these new results in Section 4.

\section{Observations}

\begin{table*}
\scriptsize
\centering
\begin{minipage}{\textwidth}
\caption{Radial velocities. \label{tbl-1}}
\centering
\begin{tabular}{ccccccccc}
\hline\hline
 HJD       & $v_{\rm DA}$ & $v_{\rm DQ}$    & HJD        & $v_{\rm DA}$ & $v_{\rm DQ}$ & HJD        & $v_{\rm DA}$ & $v_{\rm DQ}$    \\
($2450000+$) & (\kms) & (\kms) & ($2450000+$) & (\kms) & (\kms) & ($2450000+$) & (\kms) & (\kms)   \\
\hline
5507.83298 &\phantom{$-$}$152.0$ & $-65.6$ & 5930.90103 &$-$37.5             & ... & 5931.96734 &\phantom{00}$-$5.7 & ... \\
5898.10485 &   ...               & $107.1$ & 5930.95874 &\phantom{0}$-$9.3   & ... & 5932.60763 &\phantom{$-$}181.5 & ... \\
5899.19171 &\phantom{0}$-36.9$   & $114.9$ & 5931.63937 &\phantom{$-$}82.8   & ... & 5932.67638 &\phantom{$-$}155.3 & ... \\
5930.64584 &\phantom{$-$00}7.5   & ...     & 5931.69918 &\phantom{$-$}62.6   & ... & 5932.80138 &\phantom{$-$0}88.1 & ... \\
5930.71365 &\phantom{0}$-$32.9   & ...     & 5931.77189 &\phantom{$-$}11.7   & ... & 5932.91735 &\phantom{$-$0}32.4 & ... \\
5930.78652 &\phantom{0}$-$36.3   & ...     & 5931.79347 &\phantom{$-$}17.1  & ... & 5932.96735 &\phantom{$-$00}8.4 & ... \\
5930.84654 &\phantom{0}$-$24.6   & ...     & 5931.90289 &$-$30.7  & ... & 5944.85259 &     ...           & $100.6$    \\
\hline
\end{tabular}\\
\end{minipage}
\end{table*}

\begin{figure}
\centering
\includegraphics[width=1.0\columnwidth]{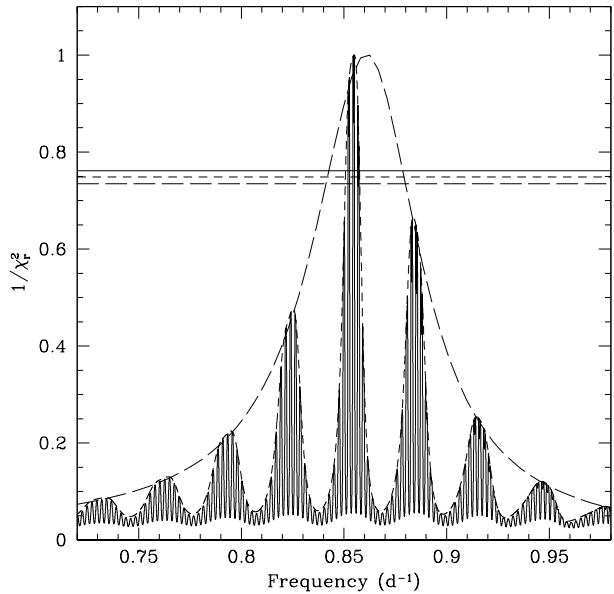}%
\caption{Frequency analysis (inverse of the reduced $\chi_r^2$ versus frequency, $1/P$) of the H$\alpha$ data ($v_{\rm DA}$, Table~\ref{tbl-1}) including
the KPNO data only (long-dashed line), the KPNO and SSO data (short-dashed line),
and the complete data set (X-shooter, KPNO, and SSO; full line). The horizontal lines
mark 1$\sigma$ confidence levels for the corresponding data sets.}
\label{fig_per}
\end{figure}

\citet{ven2012} described the discovery and initial observations of the double degenerate NLTT~16249: on UT 2010 March 26, they obtained 
three low-dispersion spectra using the R-C spectrograph attached to the 4~m telescope at Kitt Peak National Observatory (KPNO).
Next, on UT 2010 November 7, they obtained an echelle spectrum using the X-shooter spectrograph attached to the UT2 at Paranal
Observatory. 

The apparent radial velocity variations and the composite nature of the original spectra prompted us to obtain a more comprehensive
time series. We re-observed NLTT~16249 at Siding Spring Observatory (SSO), KPNO, and 
the MDM Observatory (on Kitt Peak).

On UT 2011 December 2-3, we obtained a set of spectra with the Wide Field Spectrograph
\citep[WiFeS,][]{dop2007} attached to the 2.3m telescope at SSO.
On December 2, we obtained three exposures of 15 minutes each, and, on December 3, two exposures of 30 minutes each.
We used the B3000 and R7000 gratings which provided 
spectral ranges of 3200-5900 \AA\ at a resolution $R=\lambda/\Delta\lambda = 3000$, and 5300-7000 \AA\ 
at $R = 7000$, respectively. We used the RT560 dichroic beam-splitter to 
separate the incoming light into its red and blue components. 
We maximized the signal-to-noise ratio of each observation by extracting the spectrum from the 
most significant ($\la 6$) traces. Each trace was wavelength and flux 
calibrated prior to co-addition. The spectra were wavelength 
calibrated using NeAr arc spectra that were obtained following each 
observation. 

Next, on 2012 January 4-6, we used the 4~m telescope and R-C spectrograph at KPNO. The KPC-24 grating in second order and the T2KA CCD delivered
a spectral resolution of $\sim$0.9\AA\ from 6030 to 6720\AA\ with the slit width set to 1.5$\arcsec$. We sorted the first and second orders using a
GG495 filter. We obtained 17 exposures of 30 minutes, each one followed by a comparison arc (HeNeAr). 

Finally, on 2012 January 18, we used the MDM 2.4~m Hiltner telescope and modular spectrograph (modspec).
A thick 2048$^2$ CCD gave 1.7 \AA\ pixel$^{-1}$ and $\sim3.5$ \AA\ resolution from
4460 to 7770 \AA, with vignetting toward the ends
of the range.
We obtained three consecutive exposures (5, 5, and 10 minutes).
All spectra were reduced using standard IRAF procedures. 

\section{Analysis}

\subsection{Radial velocities and period analysis}

The narrow H$\alpha$ line core proved suitable for precise radial velocity measurements 
of the DA white dwarf. When available, the sharp edges of the C$_2$ and CN band heads 
also offer means to measure relative DQ velocities. Consequently, we measured the DA radial 
velocity by fitting Gaussian functions to the narrow H$\alpha$ core ($v_{\rm DA}$ in Table~\ref{tbl-1}). 
The measurement errors were estimated by varying
the location of the continuum when fitting the line core and we noted fluctuations of 2-5 \kms\ depending on the signal-to-noise ratio.
The centering of the target on the slit is the largest source of systematic errors. These errors
were estimated
at $\sim 1/10$ of a resolution element or $\sim 5$ \kms\ for a resolution $R\approx7000$. On the other hand,
residuals in the wavelength calibration never exceeded 1-2 \kms\ and were similar to velocity fluctuations of 
the \ion{O}{1}$\lambda\lambda6300.23,6363.88$ sky lines.
Adding systematic and random errors, we conservatively estimated the total error at 10 \kms.

Next, we re-measured the DQ velocity in the UVB X-shooter spectrum using the C$_2$ Swan
band head at 5162.2\AA\ and the CN violet band head at 3883.4\AA\ and listed the
average measurement in Table~\ref{tbl-1}. The measurements are consistent with those originally employed
by \citet{ven2012}. The MDM and SSO spectra were then cross-correlated
with the X-shooter spectrum to extract new DQ velocities ($v_{\rm DQ}$ in
Table~\ref{tbl-1}). Consecutive spectroscopic series obtained at SSO or MDM were co-added
to maximise the signal-to-noise ratio and decrease the error in velocity measurements. 
The original KPNO data obtained by \citet{ven2012} are not included in the present set because of their lower spectral resolution ($R \approx 1000$).

Finally, we fitted the sinusoidal function $v=\gamma+K\times\sin{(2\pi (t-T_0)/ P)}$ 
as a function of time (HJD $t$) to the DA radial velocity
data and determined simultaneously the initial epoch ($T_0$), period ($P$), 
mean velocity ($\gamma$) and velocity
semi-amplitude ($K$).
We normalized the $\chi^2$ function so that the minimum reduced 
$\chi_{r,min}^2 \equiv 1$. The $1\sigma$ range is then estimated using
$\chi_{r,1\sigma}^2 = \chi_{r,min}^2 + c/(N-p) = 1+4.7/(N-4)$, where $N$ 
is the number of measurements, $p=4$ is the number of parameters,
and $c=4.7$ is the appropriate constant for a $1\sigma$ error range and 4 parameters.
Figure~\ref{fig_per} shows incremental frequency ($1/P$) analyses. 
The period
extracted from the KPNO radial velocity data alone ($N=17$),
\begin{displaymath}
P = 1.161\pm0.026\,{\rm d},
\end{displaymath}
is sufficiently accurate to constrain the DA velocity amplitude and mass function of the DQ companion.
Adding SSO measurements to the set ($N=18$) reduced the period uncertainty and allowed accurate phasing of the DQ velocity measurements as well
(Table~\ref{tbl-2}) with a period of
\begin{displaymath}
P = 1.1697_{-0.0037}^{+0.0057}\,{\rm d}.
\end{displaymath}
Adding X-shooter data to the set ($N=19$) resolved the solution into a triplet 
with each of the triplet members split into two possible periods. Focusing 
on the central peak, the first of the two solutions locates 
the X-shooter measurement near phase 0.15 ($P=1.1697$\,d) while the 
second locates it symmetrically about the 
quadrature and near phase 0.35 ($P=1.1703$\,d). 
Either solution is acceptable and the ambiguity does not affect the resulting 
orbital parameters for the DA and DQ stars.
The same pattern is reproduced in the side peaks of the triplet but 
with a lesser significance. 
Table~\ref{tbl-2} lists the adopted orbital parameters.  
The initial epoch $T_0$ corresponds to the lower conjunction of the DA white dwarf.
Note that the period $P$ and the initial epoch $T_0$ are strongly anti-correlated.

\begin{figure}
\centering
\includegraphics[width=1.0\columnwidth]{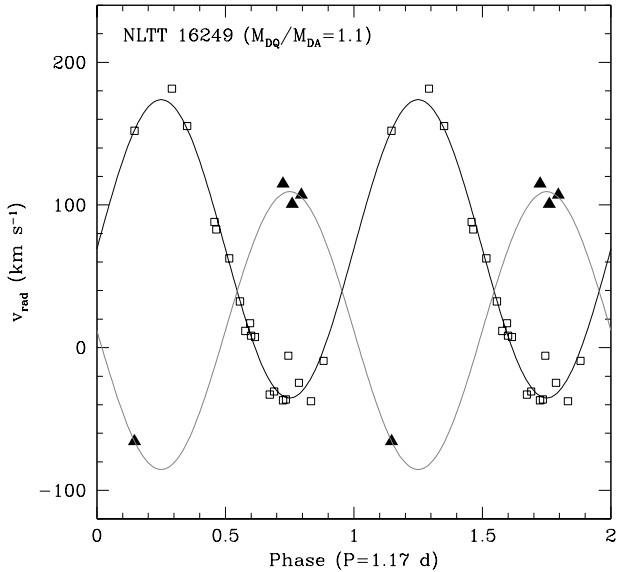}%
\caption{All radial velocity measurements from Table~\ref{tbl-1} folded on the
orbital period: DA white dwarf $v_{\rm DA}$ (open squares) and DQ white dwarf
$v_{\rm DQ}$ (full triangles). The velocity amplitudes constrain the mass ratio to
nearly unity ($q\equiv M_{\rm DQ}/M_{\rm DA} = 1.1\pm0.1$).
} 
\label{fig_DADQ}
\end{figure}

The period and velocity amplitudes constrain the total mass of the system to 
$(M_{\rm DA}+M_{\rm DQ}) \sin^3{i} = 1.01\pm0.09\,M_\odot$, where $i$ is the
inclination of the orbital plane, or $M_{\rm DA}+M_{\rm DQ}\ga 0.92\,M_\odot$.
The combined velocity amplitude is $K_{\rm DA}+K_{\rm DQ} = 202$ \kms. The same amplitude
estimated by \citet{ven2012}, $K_{\rm DA}+K_{\rm DQ}\ga 320$ \kms, was in error with
the correct constraint at only half of that value ($\ga 160$ \kms). 
The consequences for the merger time-scale are discussed in the following Section.

Figure~\ref{fig_DADQ} shows the radial velocity phased with the orbital period. The 
X-shooter data are phased by adopting one of the two possible aliases ($P=1.1697$ d). The dispersion in H$\alpha$ velocity measurements 
was only 10 \kms\ and consistent with the estimated velocity errors. The DQ velocity curve is offset by $-60$ \kms\ relative to the DA velocity 
curve. The effect cannot be attributed to a difference in 
gravitational redshifts because the mass ratio is close to unity. Instead, the pressure shift
of molecular band positions is a likely cause for this effect. In the following we utilize
the DA mean-velocity in the calculation of the systemic velocity and Galactic kinematics.

\begin{figure}
\centering
\includegraphics[width=1.0\columnwidth]{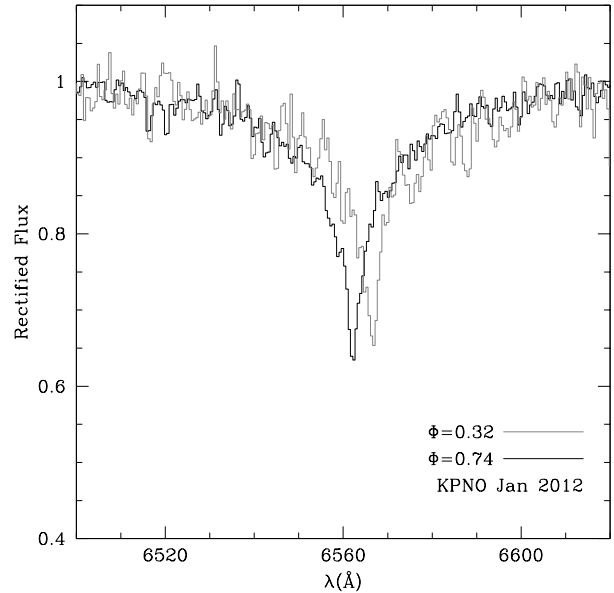}%
\caption{Sets of co-added H$\alpha$ spectra near quadratures. The H$\alpha$ line core shifted by
$\sim4$\AA\ between the two sets.
} 
\label{fig_alpha}
\end{figure}

Figure~\ref{fig_alpha} shows two sets of co-added KPNO spectra, one near each quadrature.
The velocity offset ($\Delta v\approx 180$ \kms) proved readily measurable.

If we adopt the spectroscopic mass estimate for the DA white dwarf based on the measured (\teff,$\log{g}$) from \citet{ven2012},
$M_{\rm DA}=0.829\pm0.096\,M_\odot$ \citep[following the mass-radius relations of][]{ben1999}, and the orbital mass ratio ($q=1.1\pm0.1$), then the total mass of the system is in the
range 1.466-2.035$\,M_\odot$. Consequently, the mass function constrains the inclination
to $i=52-70^\circ$, and the possibility of eclipses is ruled out.

\begin{table}
\centering
\begin{minipage}{\columnwidth}
\caption{Binary parameters. \label{tbl-2}}
\centering
\begin{tabular}{cc}
\hline\hline
Parameter  & Measurement \\
\hline
Period     & $P=1.1697_{-0.0037}^{+0.0057}$ d \\
Initial epoch (HJD) & $T_0=2455898.34_{+0.10}^{-0.15}$ \\
Semi-amplitude (DA) & $K_{\rm DA} = 104.6\pm3.7$ \kms \\
Mean velocity (DA) & $\gamma_{\rm DA} = 69.3\pm2.8$ \kms \\
Semi-amplitude (DQ) & $K_{\rm DQ} = 97.4\pm5.1$ \kms \\
Mass ratio         & $q\equiv M_{\rm DQ}/M_{\rm DA} = 1.1\pm0.1$ \\
\hline
\end{tabular}\\
\end{minipage}
\end{table}

\subsection{Age, evolution, and kinematics}

The cooling age of the DA white dwarf is estimated at $\tau_{\rm cool} = 1.9\pm0.5$ Gyr.
A high mass for both white dwarfs implies relatively massive, $M_i\ga 3 \,M_\odot$ \citep[see a discussion in][]{fer2005}, 
hence short-lived progenitors \citep[$\sim0.48$ Gyr,][]{sch1992}, constraining the
total age of the system to less than 3 Gyr.

Using our measured period (1.17 d) and mass ratio $q=1.1\pm0.1$, and constraining the DA
mass to $0.829\pm0.096\,M_\odot$, 
the merger time is estimated at $10^{11}$~yr following \citet{rit1986}. Therefore, the system is not 
a candidate type 1a supernova progenitor.

The systemic velocity was obtained by subtracting the calculated gravitational redshift ($\gamma_{\rm grav}=54\pm6$ \kms) from the
mean velocity of the DA white dwarf ($\gamma_{\rm DA}=69\pm3$ \kms), $\gamma_{\rm sys} = \gamma_{\rm DA}-\gamma{\rm grav} = 15\pm7$ \kms.
Adopting a distance of $40\pm6$ pc, the systemic velocity and proper-motion measurements \citep[$\mu_\alpha\cos{\delta}=2\pm6,\ \mu_\delta=-140\pm6$ mas yr$^{-1}$,][]{sal2003}
correspond to the Galactic velocity components $(U,V,W)=(0\pm8,-27\pm6,-6\pm2)$ \kms\ following \citet{joh1987}. The system belongs to the thin-disk
population \citep{pau2006} in agreement with its young age ($\la 3$ Gyr).

\section{Summary and Discussion}

We have shown that the peculiar DQ white dwarf in NLTT~16249 is in a close 1.17 d orbit with a
DA white dwarf companion. Close double-degenerate stars are common and \citet{max1999} reported
a fraction of 5 to 19\% of close pairs in their radial velocity survey.  
White dwarfs with composite spectra indicative of DA plus DB (He{\sc i}) or 
DA plus DC (cool He-rich) pairs are also relatively common in large surveys \citep[see, e.g.,][]{lim2010,tre2011}. However, only two DA+DQ pairs are known: NLTT~16249
and SDSS~J153210.04+135615.0 (NLTT~40489) which was recently identified by \citet{gia2012}. 

The velocity amplitudes in NLTT~16249 imply that the DQ is slightly more
massive than its companion in which case it is more likely that the DQ formed first, both
from $\sim3\,M_\odot$ progenitors, unless
stable mass transfer reversed the initial mass ratio \citep[for detailed scenarios see][]{nel2001}.
The relatively long orbital period reported here rules out the prospect of a merger 
within a Hubble time.

The DQ white dwarf is peculiar and the nitrogen concentration in its atmosphere is, so far, unique.
\citet{alt2005} followed the evolution of a 2.7$M_\odot$ main-sequence star past the 
PG1159 spectroscopic stage and down the cooling track up to the DQ stage (0.6$M_\odot$).
The star follows a ``born-again'' He-flash loop that resets its evolution onto the post-asymptotic
giant-branch (post-AGB), considerably modifying the chemical structure of the outer layers that
an otherwise normal post-AGB would possess.
In particular, \citet{alt2005} predicted measurable concentrations of nitrogen and oxygen
in the atmosphere of DQ descendents of born-again stars. 

The nitrogen abundance measured in the DQ component
of NLTT~16249 is qualitatively similar to the abundance predicted for the 10,500~K model,
the lowest temperature considered by \citet{alt2005}. However, the ${\rm C/N}$ ratio for that
particular model is close to 300, i.e, much larger than measured in NLTT~16249 (${\rm C/N}\approx 50$). In the born-again context, the difference could be attributed to a different 
progenitor mass which could affect the outcome of the chemical evolution, although
such models are not available to us. \citet{alt2005} also predicted a concentration of
oxygen between that of carbon and nitrogen (${\rm C/O}\approx 100$). 
It is possible  to constrain the ${\rm C/O}$ abundance ratio in DQ white dwarfs
using the near and far ultraviolet CO ``Fourth-Positive" (4P) bands 
along with CN violet and C$_2$ Swan bands.
Finally, the $^{13}{\rm C}/^{12}{\rm C}$ isotopic ratio expected for the same evolutionary track
is well below detection limits reported by \citet{ven2012} in the case of the DQ in NLTT~16249.

What would be the role, if any, played by binary interactions in shaping the 
chemical structure of the DQ progenitor in NLTT~16249 ? 
The DQ white dwarfs are not known to show nitrogen in their spectra \citep[see][]{duf2005}. 
Therefore, it may not be a coincidence that the only known detection of
nitrogen in a DQ white dwarf is also that of a white dwarf in a close double-degenerate
system. 
In that context as well, it remains to be shown that detectable traces of nitrogen
would be the expected outcome of mass-transfer or early envelope ejection.

\acknowledgments

S.V. and A.K. acknowledge support from 
the Grant Agency of the Czech Republic (GA \v{C}R P209/10/0967, GA \v{C}R P209/12/0217). 
This work was also supported by the project RVO:67985815 in the Czech Republic.
We thank Donna Burton for invaluable help during the December observations at Siding Spring Observatory.
J.R.T. gratefully acknowledges support from NSF grants AST-0708810 and AST-1008217.


\begin{thebibliography}{}

\bibitem[\protect\citeauthoryear{Althaus et al.}{2005}]{alt2005} Althaus, L.~G., Serenelli, A.~M., Panei, J.~A., C{\'o}rsico, A.~H., Garc{\'{\i}}a-Berro, E., Sc{\'o}ccola, C.~G., 2005 A\&A, 435, 631 
\bibitem[\protect\citeauthoryear{Benvenuto 
\& Althaus}{1999}]{ben1999} Benvenuto, O.~G., Althaus, L.~G. 1999, MNRAS, 303, 30 
\bibitem[\protect\citeauthoryear{Dopita et al.}{2007}]{dop2007} Dopita, M., Hart, J, McGregor, P., Oates, P., Jones, D. 2007, Ap\&SS, 310, 255
\bibitem[\protect\citeauthoryear{Dreizler 
\& Heber}{1998}]{dre1998} Dreizler, S., Heber, U. 1998, A\&A, 334, 618 
\bibitem[\protect\citeauthoryear{Dufour, Bergeron, 
\& Fontaine}{2005}]{duf2005} Dufour, P., Bergeron, P., Fontaine, G. 2005, ApJ, 627, 404 
\bibitem[\protect\citeauthoryear{Ferrario et al.}{2005}]{fer2005} Ferrario, L., Wickramasinghe, D., Liebert, 
J., Williams, K.~A. 2005, MNRAS, 361, 1131 
\bibitem[\protect\citeauthoryear{Giammichele et al.}{2012}]{gia2012} Giammichele, N., Bergeron, P., Dufour, P. 2012, ApJS, 199, 29
\bibitem[\protect\citeauthoryear{Johnson \& Soderblom}{1987}]{joh1987} Johnson, D.~R.~H., Soderblom, D.~R. 1987, AJ, 93, 864 
\bibitem[\protect\citeauthoryear{Koester \& Knist}{2006}]{koe2006} Koester, D., Knist, S. 2006, A\&A, 454, 951 
\bibitem[\protect\citeauthoryear{Limoges \& Bergeron}{2010}]{lim2010} Limoges, M.-M., Bergeron, P. 2010, ApJ, 714, 1037 
\bibitem[\protect\citeauthoryear{MacDonald, Hernanz, \& Jose}{1998}]{mac1998} MacDonald, J., Hernanz, M., Jose, J. 1998, MNRAS, 296, 523
\bibitem[\protect\citeauthoryear{Maxted \& Marsh}{1999}]{max1999} Maxted, P.~F.~L., Marsh, T.~R. 1999, MNRAS, 307, 122 
\bibitem[\protect\citeauthoryear{Nelemans et 
al.}{2001}]{nel2001} Nelemans, G., Yungelson, L.~R., Portegies Zwart, S.~F., Verbunt, F. 2001, A\&A, 365, 491
\bibitem[\protect\citeauthoryear{Pauli et al.}{2006}]{pau2006} Pauli, E.-M., Napiwotzki, R., Heber, U., Altmann, M., Odenkirchen, M. 2006, A\&A, 447, 173 
\bibitem[\protect\citeauthoryear{Pelletier et al.}{1986}]{pel1986} Pelletier, C., Fontaine, G., Wesemael, F., Michaud, G., Wegner, G. 1986, ApJ, 307, 242 
\bibitem[\protect\citeauthoryear{Ritter}{1986}]{rit1986} Ritter, H. 1986, A\&A, 169, 139 
\bibitem[\protect\citeauthoryear{Salim \& Gould}{2003}]{sal2003} Salim, S., Gould, A. 2003, ApJ, 582, 1011 
\bibitem[\protect\citeauthoryear{Schaller et al.}{1992}]{sch1992} Schaller, G., Schaerer, D., Meynet, G., Maeder, A. 1992, A\&AS, 96, 269 
\bibitem[\protect\citeauthoryear{Tremblay, Bergeron, \& Gianninas}{2011}]{tre2011} Tremblay, P.-E., Bergeron, P., Gianninas, A. 2011, ApJ, 730, 128 
\bibitem[Vennes \& Kawka(2012)]{ven2012} Vennes, S., \& Kawka, A.\ 2012, ApJL, 745, L12 
\bibitem[\protect\citeauthoryear{Werner, Heber, \& Hunger}{1991}]{wer1991} Werner, K., Heber, U., Hunger, K. 1991, A\&A, 244, 437 
\bibitem[Werner \& Herwig(2006)]{wer2006} Werner, K., \& Herwig, F.\ 2006, \pasp, 118, 183

\end{thebibliography}
\end{document}